\documentclass[dvipsnames]{osa-article}
\usepackage{xcolor}
\usepackage{graphicx}
\usepackage{bm}
\usepackage{siunitx}
\journal{oe}

\usepackage[textsize=scriptsize]{todonotes}
\usepackage{xargs}
\newcommandx{\vnm}[2][1=]{\todo[color=blue!30, #1]{vnm: #2}}
\newcommandx{\bdl}[2][1=]{\todo[color=Emerald!30, #1]{bdl: #2}}

\begin{document}

\title{Large depth-of-field tracking of colloidal spheres in holographic
microscopy by modeling the objective lens}


\author{Brian Leahy\authormark{1, 3}, Ronald Alexander\authormark{2, 3},
Caroline Martin\authormark{1}, Solomon Barkley\authormark{2}, and
Vinothan Manoharan\authormark{1, 2, *}}

\address{
\authormark{1}Harvard John A. Paulson School of Engineering and Applied
Sciences, Harvard University, Cambridge MA 02138, USA\\
\authormark{2}Department of Physics, Harvard University, Cambridge MA 02138, USA\\
\authormark{3}{These authors contributed equally to this work.}
}

\email{\authormark{*}vnm@seas.harvard.edu}

\begin{abstract}
  Holographic microscopy has developed into a powerful tool for 3D
  particle tracking, yielding nanometer-scale precision at high frame
  rates. However, current particle tracking algorithms ignore the effect
  of the microscope objective on the formation of the recorded hologram.
  As a result, particle tracking in holographic microscopy is currently
  limited to particles well above the microscope focus. Here, we show
  that modeling the effect of an aberration-free lens allows tracking of
  particles above, near, and below the focal plane in holographic
  microscopy, doubling the depth of field. Finally, we use our model to
  determine the conditions under which ignoring the effect of the lens
  is justified, and in what conditions it leads to systematic errors.
\end{abstract}

\renewcommand{\vec}[1]{\mathbf{#1}}
\newcommand{\unit}[1]{\mathbf{\hat{#1}}}
\newcommand{\tensor}[1]{\bm{#1}}

\renewcommand{\Im}{\mathrm{Im}}
\renewcommand{\Re}{\mathrm{Re}}

\newcommand{\hologram}{H}

\newcommand{\micron}{\SI{}{\micro\meter}}
\newcommand{\microliter}{\SI{}{\micro\liter}}
\newcommand{\xp}{\vec{x}_\mathrm{p}}
\newcommand{\rp}{r_\mathrm{p}}
\newcommand{\phip}{\phi_\mathrm{p}}
\newcommand{\thetap}{\theta_\mathrm{p}}
\newcommand{\zp}{z_\mathrm{p}}
\newcommand{\rhop}{\rho_\mathrm{p}}

\newcommand{\So}{S_\mathrm{o}}
\newcommand{\rado}{d_\mathrm{o}}
\newcommand{\thetao}{\theta_\mathrm{o}}
\newcommand{\phio}{\phi_\mathrm{o}}
\newcommand{\Si}{S_\mathrm{i}}
\newcommand{\dOo}{d\Omega_\mathrm{o}}
\newcommand{\radi}{d_\mathrm{i}}
\newcommand{\thetai}{\theta_\mathrm{i}}
\newcommand{\phii}{\phi_\mathrm{i}}
\newcommand{\dOi}{d\Omega_\mathrm{i}}

\newcommand{\rhod}{\rho_\mathrm{det}}
\newcommand{\phid}{\phi_\mathrm{det}}

\newcommand{\wavefront}{\Phi}
\newcommand{\aberration}{\Phi(\thetao, \phio)}

\newcommand{\Esc}{\vec{E}_\mathrm{sc}}
\newcommand{\Ein}{\vec{E}_\mathrm{in}}
\newcommand{\Edet}{\vec{E}_\mathrm{sc,det}}
\newcommand{\Eindet}{\vec{E}_\mathrm{in,det}}
\newcommand{\Eindetconj}{\vec{E}^*_\mathrm{in,det}}

\newcommand{\farfield}{\tensor{S}}

\newcommand{\Izero}{\mathcal{I}_0}
\newcommand{\Itwo}{\mathcal{I}_2}

\newcommand{\lensangle}{\beta}

\newcommand{\crosssection}{\Sigma}

\bibliography{full_bibliography}

\section{Introduction}

\begin{figure}
  \centering
  \includegraphics{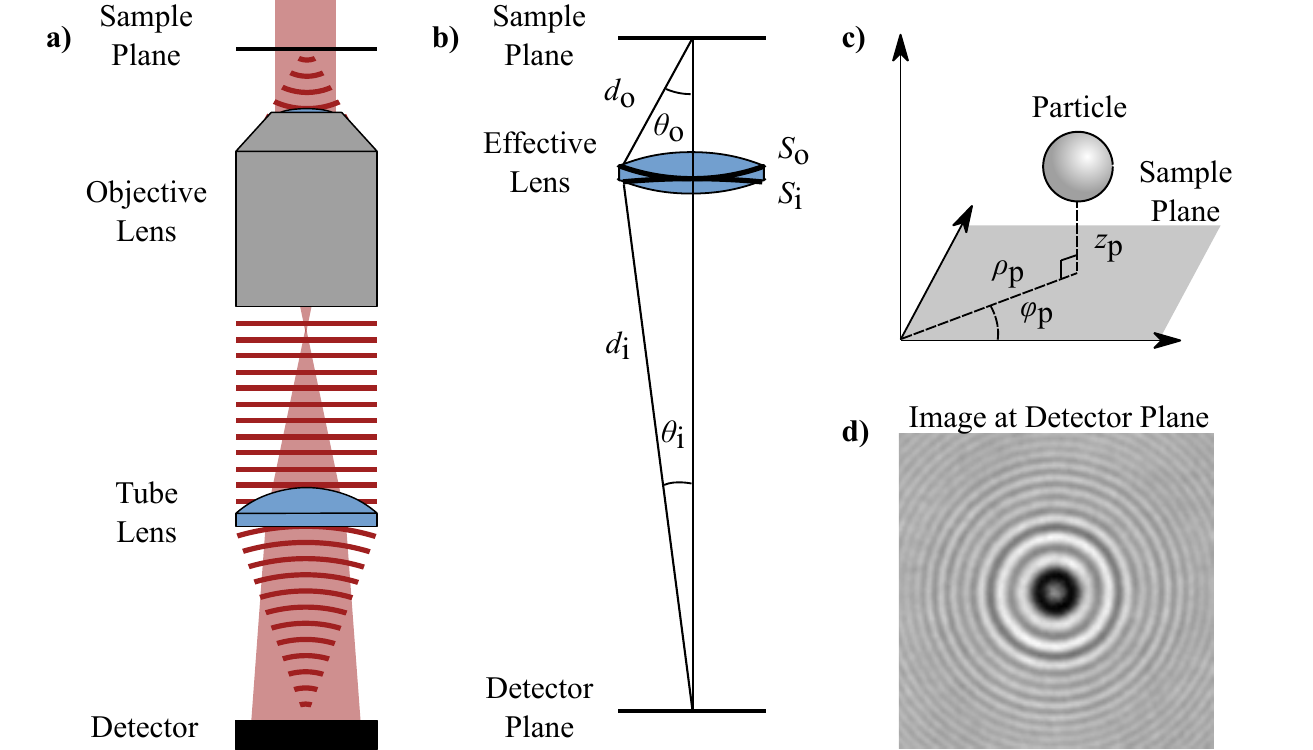}
  \caption{(a) In in-line digital holographic microscopy, a collimated
    laser (light red) illuminates a sample, which scatters light (dark
    red wavefronts). An objective collects the transmitted and some of
    the scattered light, and a tube lens focuses this light onto a
    digital camera. (b) We treat the objective and tube lens combination
    as a single effective lens. The thick lines illustrate the Gaussian
    reference spheres $\So$ and $\Si$. The thin lines illustrate the
    coordinates $(\rado, \thetao)$ and $(\radi, \thetai)$ in one plane;
    the coordinates $\phio$ and $\phii$ describe the rotation about the
    optical axis. (c) The coordinate system for the particle relative to
    the objective's focus. (d) A recorded hologram of a polystyrene
    sphere.}
  \label{fig:lensmodel}
\end{figure}

Digital holographic microscopy excels at fast, precise three-dimensional
(3D) imaging of colloidal particles and nanoparticles. In a typical
in-line digital holographic microscope, coherent light illuminates a
sample, an objective collects the scattered and transmitted light, and a
digital camera records the resulting interference pattern, or hologram
(Fig.~\ref{fig:lensmodel}a). Compared to bright-field and confocal
microscopy, holographic microscopy has three advantages for imaging
small particles. First, because a hologram records both the phase and
amplitude of the scattered wave, a single camera exposure captures
information about the particles' 3D position, shape, and size. Second,
because the technique does not require fluorescent labels, the incident
beam power is not limited by bleaching. Third, the depth of field in
holography is tens to hundreds of micrometers. Combined, these three
advantages give holographic microscopy an enormous dynamic range: 3D
information can be captured at rates of thousands of frames per second
and over durations of hours or more. Moreover, the precision of the
technique is unparalleled: micrometer-sized particles can be localized
to nanometer-scale precision in all three dimensions~\cite{Lee2007}.
Holographic microscopes have been used to track both individual
colloidal particles~\cite{Sheng2006, Fung2011measuring} and colloidal
clusters~\cite{Fung2013} in 3D, to watch colloidal particles breach
interfaces~\cite{Kaz2011}, and to characterize 3D fluid
turbulence~\cite{hinsch2002holographic, garcia2008}. Because it is
noninvasive, holographic microscopy has also proven useful for imaging
living specimens, including bacterial~\cite{Wang2016} and eukaryotic
cells~\cite{khmaladze2008phase, choi2009, kemper2011, marquet2005}.

Extracting 3D information about the sample, such as the 3D position of a
particle, requires analyzing the recorded hologram. One method, based on
Gabor's original conception of holography~\cite{Gabor1948new}, is to
reconstruct the electric field everywhere in space using the phase and
amplitude information recorded in the hologram and to identify the
particle's location from the reconstructed field~\cite{schnars2002}.
This approach is indirect: the hologram is processed to yield a
reconstruction of the field, which is then processed again to extract
the particle's position. Furthermore, the depth resolution is
poor~\cite{Garcia-Sucerquia2006}, and artifacts arise when the particle
is comparable in size to the wavelength of the incident
light~\cite{Pu2003}.

A more direct and precise approach is forward modeling combined with
inference, wherein the recorded hologram is compared to a simulation
from a scattering model~\cite{ovryn2000imaging, Lee2007,
cheong2010strategies}. The parameters of the model---which might include
the particle's position, size, and refractive index---can then be
determined by fitting the simulated hologram to the measured one. With
this approach, one can not only track particles with high
precision~\cite{Kaz2011, krishnatreya2014measuring, obrien2019above} but
also calculate uncertainties on the particle positions using Bayesian
inference~\cite{dimiduk2016bayesian}.

Although many forward models start from exact solutions of Maxwell's
equations to describe the particle scattering~\cite{Fung2011measuring,
Fung2012imaging, Wang2014using}, modeling the hologram formation from
the scattered wave involves approximations that ultimately limit our
ability to track particles. In most approaches, the hologram
$\hologram{}(\xp{})$ recorded on the detector is modeled as the
magnified image of the scattered field $\Esc{}$ and the incident field
$\Ein{}$ at the focal plane of the objective:
\begin{equation}
\hologram{}(\xp{}) = \left|\Ein{} + \alpha \Esc{}(\xp{}) \right|^2,
\label{eqn:lenslessmodel}
\end{equation}
where $\xp$ is the particle's position relative to the point in the
focal plane conjugate to a detector pixel, and $\alpha$ is a
field-rescaling parameter that accounts for experimental imperfections
in the optical setup~\cite{Lee2007}. When the particle is well above the
focal plane, this \emph{lensless model} works well. However, when the
particle is near or below the focal plane, the lensless model makes the
unphysical prediction that the hologram results from the internal or
backscattered electric field. Accordingly, the lensless model is not
useful for tracking the motion of particles near or below the focus of
the objective; therefore, this approximation limits the depth of
field. Furthermore, the lensless model does not account for physical
effects of the lens such as the Gouy phase shift~\cite{Feng2001,
atilgan2011reflectivity, Mertz2010introduction, Wilson20123d}. As we
shall show, modeling these effects is critical to accurately simulating
the hologram when the particle is near the focus.

Here we show that explicitly modeling the effects of the objective lens
doubles the depth of field of the holographic microscope by reducing
systematic errors near and below the objective's focus. We model the
lens by describing the scattered electric field as defocused,
forward-scattered light, in an approach similar to that of Ovryn and
Izen~\cite{ovryn2000imaging} but with additional simplifications that
allow for a compact representation of the recorded hologram. With this
model, we can track particles above, near, and below the focus of the
objective. Finally, by comparing the results of our model to the
lensless model, we determine the conditions under which the lens can be
safely ignored.

\section{Theory}

To create an analytically tractable model for the effect of the lens, we
simplify the full optical train of an inline holographic microscope.
First, we ignore any reflections or aberrations due to the optical
interface of the cover slip~\cite{Hell1993, ovryn2000imaging}, because
these can be minimized with an immersion lens. Second, we describe the
combination of the objective and tube lens as a single effective lens
with the same magnification and numerical aperture, because this is the
simplest model that captures the phase, polarization, and resolution
effects of the imaging system~\cite{Richards1959}. Third, we assume an
aberration-free and translationally invariant aplanatic system, which is
a good approximation for modern microscope lenses. This simplified model
of the imaging system is shown in Fig.~\ref{fig:lensmodel}b.

To calculate the hologram on the detector plane, we treat
the scattered beam and the incident beam separately. We calculate the
image of the scattered field following derivations of microscope
point-spread functions~\cite{Richards1959,Visser1991,Hell1993}. In
Sec.~\ref{sec:entrance_pupil}, we evaluate the scattered electric field
on the lens's entrance pupil, represented as a Gaussian reference sphere
$\So$ centered on the object point of the effective lens. In
Sec.~\ref{sec:lens}, we model how the lens transforms the field on its
entrance pupil to that on its exit pupil, represented as the Gaussian
reference sphere $\Si$ centered on the image point of the effective
lens. In Sec.~\ref{sec:detector}, we use a diffraction integral to
propagate the scattered field from the surface $\Si$ to the detector
plane. This step yields a two-dimensional integral equation for the
field scattered from a generic particle as imaged through a
large-aperture microscope objective. In Sec.~\ref{sec:mie}, we simplify
this integral for spherically symmetric scatterers, reducing it to a
one-dimensional integral that is easily evaluated numerically. Finally,
in Sec.~\ref{sec:hologram}, we model the hologram by calculating the
interference between the scattered field and the transmitted field,
which propagates through the same optical train.

We work in physical limits relevant for holographic microscopy. We
assume that the microscope dimensions are large compared to the
wavelength of light ($k \rado{} \gg 1$, $k \radi{} \gg 1$, where $k$ is
the wavevector of the light in the medium; that the position of the
particle $\xp$ is close to the focus of the imaging system ($|\xp| /
\rado{} \ll 1$); and that the ratio of the microscope's numerical
aperture $\mathrm{NA}$ to its magnification $M$ is small---an assumption
equivalent to $\theta_\mathrm{i} \ll 1$, where $\theta_\mathrm{i}$ is
defined in Fig.~\ref{fig:lensmodel}b. For a 100$\times$ magnification,
$\mathrm{NA}=1.0$ microscope with a working distance of 0.2 mm and a 200
mm tube lens that images a particle 5 \micron{} from its focal plane
using 660 nm red light, $k \rado{} \approx 2 \times 10^3$, $k\radi{}
\approx 2 \times 10^6$, $|\xp| / \rado{} \approx 0.02$, and $\mathrm{NA}
/ M \approx 0.01$, and these approximations are well justified. We allow
the numerical aperture to be large and do not make a paraxial
approximation.

\subsection{Scattered field on the entrance pupil of the lens}
\label{sec:entrance_pupil}

We first evaluate the electric field on the Gaussian reference sphere
$\So$, as scattered from a particle located at a position $\xp$ from the
center of $\So$ (see Fig.~\ref{fig:lensmodel}b for a coordinate
diagram). Let $\vec{d}_\mathrm{o}$ be a point on $\So$, which we
represent in spherical polar coordinates from the center of $\So$ as
$(\rado{}, \thetao{}, \phio{})$. The point $\vec{d}_\mathrm{o}$ is
located at $\vec{r}_\mathrm{op} \equiv \vec{d}_\mathrm{o} - \xp$
relative to the particle. Because the lens is a macroscopic distance
from the particle, the magnitude of $\vec{r}_\mathrm{op}$ is much larger
than the wavelength of light, and the scattered electric field on
$\So{}$ is in the far-field limit. For a generic scatterer excited by
an incoming wave $E_0 \unit{x} e^{ikz}$, the scattered electric field in
the far-field limit takes the form~\cite{Hulst1981} 
\begin{equation}
\vec{E}_{\So} = \frac {E_0} {i kr_\mathrm{op}} e^{-ik{r_\mathrm{op} + ik\zp{}}} \, \farfield{}(\theta_\mathrm{op}, \phi_\mathrm{op}) \cdot \unit{x},
\label{eqn:farfield_particle}
\end{equation}
where $(r_\mathrm{op}, \theta_\mathrm{op}, \phi_\mathrm{op})$ is the
decomposition of $\vec{r}_\mathrm{op}$ in spherical polar coordinates
and $\farfield{}$ is the far-field scattering matrix. To lowest order in
$|\xp| / \rado{}$, these components are $ r_\mathrm{op} = \rado{} - {\xp
\cdot \vec{d}_o} / {\rado{}}$, $\theta_\mathrm{op} = \thetao{}$, and
$\phi_\mathrm{op} = \phio{}$, and the scattered field on the surface
$\So$ is
\begin{equation}
\vec{E}_{\So} (\rado{}, \thetao{}, \phio{})=
    \frac {E_0} {i k\rado{}} e^{-ik\rado{}}
    e^{ik \rhop{} \sin \thetao{} \cos(\phio{} - \phip{})}
    e^{ik \zp{} (1 - \cos \thetao{})} \, \farfield{}(\thetao{}, \phio{}) \cdot \unit{x},
\label{eqn:farfield_so}
\end{equation}
where $(\rhop{}, \phip{}, \zp{})$ is the particle position in
cylindrical coordinates, with $+\zp{}$ oriented away from the lens. The
paraxial approximation differs from Eq.~\eqref{eqn:farfield_so} by
making the approximations $\sin \theta_\mathrm{o} \approx
\theta_\mathrm{o}$ and $\cos \theta_\mathrm{o} \approx 1$. We do not use
the paraxial approximation because the scattered field in this
approximation does not provide any information about the particle's axial
position $\zp{}$.

\subsection{Transformation of the scattered field by the lens}
\label{sec:lens}

The lens transforms the phase, polarization direction, and magnitude of
the scattered field on $\So$ to new values on $\Si$. To understand this
transformation, we consider a simple model of a perfect
lens~\cite{Richards1959, Visser1991}. A perfect lens stigmatically
images a point at the object plane to a point on the image plane; all
rays from the object point which enter the lens pupil intersect at the
image point. By Fermat's principle, each of these rays traverses an
equal optical path. Since the surface $\So$ is a sphere centered at the
object point, each ray leaving the object point accumulates the same
phase at $\So$. Likewise, each ray leaving the second spherical surface
$\Si$ accumulates the same phase at the image point. Thus, to
stigmatically image the object point to the image point, the lens maps
the rays on the first surface $\So$ to the second surface $\Si$ with the
same constant phase shift $\wavefront{}$ for each ray.

The lens also rotates the electric field's polarization, applying the
same rotation to the polarization vectors as it does to the ray
directions~\cite{Born1983}. The lens rotates the polarization projection
along the unit vector $\unit{\theta}_\mathrm{o}$ on $\So$ to along
$\unit{\theta}_\mathrm{i}$ on $\Si$, and from $\unit{\phi}_\mathrm{o}$
to $\unit{\phi}_\mathrm{i}$, where $(\thetai{}, \phii{})$ are
coordinates on the surface $\Si{}$, and $(\unit{\theta}_\mathrm{i},
\unit{\phi}_\mathrm{i})$ are the associated unit vectors. For a typical
microscope imaging configuration, $\radi{} \gg \rado{}$, and the
polarization vectors exiting the lens will be approximately parallel to
the detector plane---that is, $\unit{\theta}_\mathrm{i} \approx \unit{x}
\cos \phi_\mathrm{i} + \unit{y} \sin \phi_\mathrm{i}$ and
$\unit{\phi}_\mathrm{i} \approx -\unit{x} \sin \phi_\mathrm{i} +
\unit{y} \cos \phi_\mathrm{i}$.

Finally, the lens slightly rescales the magnitude of the electric field
as the rays propagate from $\So$ to $\Si$~\cite{Richards1959}. Consider
the flux of energy from the rays that enter through a small surface area
$d\So$ centered at $(\thetao{}, \phio{})$ on the entrance pupil and exit
through the area $d\Si$ on the exit pupil. By conservation of energy,
the incident flux $|E_{\So}|^2 / c \times d\So$ must equal the outgoing
flux $|E_{\Si}|^2 / c \times d\Si$. For a translationally-invariant
aplanatic system of magnification $M$, the Abbe sine condition, $\sin
\thetao{} = M \sin \thetai{}$, relates the two elements of area as $\cos
\thetai{} d\Si = \cos \thetao{} d\So$. Substituting this relation into
the outgoing flux and approximating $\cos \thetai{} \approx 1$ relates
the field magnitudes as $|\vec{E}_{\Si}| =|\vec{E}_{\So}| / \sqrt{\cos
  \thetao{}}$.

Combining the phase shift, polarization rotation, and the change in
magnitude due to the lens yields the field on the exit pupil $\Si$:
\begin{equation}
\begin{aligned}
\vec{E}_{\Si}(\thetai{}, \phi_\mathrm{i}) = e^{i\wavefront{}} \frac 1 {\sqrt{\cos \thetao{}}} \Big[ &
    \left(\unit{\theta}_\mathrm{o} \cdot \vec{E}_{\So}(\thetao{}, \phio{}) \right)
        \left( \unit{x} \cos \phi_\mathrm{i} + \unit{y} \sin \phi_\mathrm{i} \right) +
    \\ &
    \left( \unit{\phi}_\mathrm{o} \cdot \vec{E}_{\So}(\thetao{}, \phio{}) \right)
        \left(-\unit{x} \sin \phi_\mathrm{i} + \unit{y} \cos \phi_\mathrm{i} \right)
\Big].
\end{aligned}
\label{eqn:field_si}
\end{equation}

\subsection{Propagation of the scattered field to detector plane}
\label{sec:detector}

Finally, the electric field propagates from $\Si$ to the detector. For a
translationally-invariant optical system, we need only consider the
electric field at the detector's center, corresponding to the center of
the sphere $\Si{}$, because examining a different location on the
detector is equivalent to shifting the particle. A Kirchoff diffraction
integral over the surface $\Si$ yields the scattered field at the center
of the detector plane:
\begin{equation}
\Edet{} = \frac {ik \radi{}} {4\pi} {e^{-ik\radi{}}} \int (1 + \cos \thetai{}) \vec{E}_{\Si} \, \dOi{},
\label{eqn:diffraction_focus_si}
\end{equation}
where $\dOi{}$ is the element of solid angle on $\Si$. Here, the Green's
function $ik / 4 \pi r \times e^{-ikr}$ is constant, because every point
on the spherical cap $\Si$ is a distance $\radi{}$ from the focus of the
imaging system. Substituting Eqs.~\eqref{eqn:farfield_so}
and~\eqref{eqn:field_si} into Eq.~\eqref{eqn:diffraction_focus_si},
approximating $\cos \thetai{} \approx 1$, and transforming the domain of
integration from $\Si$ to $\So$ using the Abbe sine condition for the
Jacobian $\dOi{} / \dOo{}$ yields the scattered electric field at the
detector:
\begin{equation}
\begin{aligned}
\Edet{} = & \frac 1 {2\pi} \frac {E_0} {M} e^{-ik\radi{}} e^{i\wavefront{}} e^{-ik\rado{}} \times \\ &
    \int_{\phio{}=0}^{2\pi} \int_{\thetao{}=0}^{\lensangle} \,
         e^{ik\rhop{} \sin \thetao{} \cos(\phio{} - \phip{})} e^{ik \zp{}(1 - \cos \thetao{})} \times
        \\ & \quad \quad
        \Big[ \left(\unit{\theta}_\mathrm{o} \cdot
        \farfield{}(\thetao{}, \phio{}) \cdot \unit{x} \right)
            \left( \unit{x} \cos \phio + \unit{y} \sin \phio \right) +
        \\ & \quad \quad
        \left( \unit{\phi}_\mathrm{o} \cdot \farfield{}(\thetao{}, \phio{})\cdot \unit{x} \right)
            \left(-\unit{x} \sin \phio + \unit{y} \cos \phio \right)
    \Big]
    \sqrt{\cos \thetao{}} \, \sin \thetao{} \, d\thetao{} d\phio{},
\end{aligned}
\label{eqn:generic_scatterer_lens}
\end{equation}
where $\lensangle$ is the acceptance angle of the objective, related to
the numerical aperture $\mathrm{NA}$ through the immersion fluid index
$n_f$ as $\mathrm{NA} = n_f \sin \lensangle{}$.
Equation~\eqref{eqn:generic_scatterer_lens} gives the electric field
from an arbitrary scatterer as imaged on the detection plane of a
microscope.

\subsection{Simplified form for Mie scatterers}
\label{sec:mie}

Up to this point, we have made no assumptions about the type of
scatterer. For spherically symmetric scatterers, the symmetry of the
scattered field greatly simplifies
Eq.~\eqref{eqn:generic_scatterer_lens}, allowing us to analytically
integrate over the $\phi$ coordinate. The far-field scattering matrix
becomes
\begin{equation}
\farfield{}(\thetao{}, \phio{}) \cdot \unit{x} = S_\parallel (\thetao{}) \cos(\phio{}) \unit{\theta}_\mathrm{o} - S_\perp (\thetao{}) \sin(\phio{}) \unit{\phi}_\mathrm{o},
\label{eqn:miescattering}
\end{equation}
where $S_\parallel$ and $S_\perp$ are given by Mie
theory~\cite{Hulst1981}. Substituting the scattered field into
Eq.~\eqref{eqn:generic_scatterer_lens} yields integrals over $\phio{}$
of the form $\int_0^{2\pi} \cos(n\phio{}) e^{ix\cos\phio{}} \, d\phio{}$
and $\int_0^{2\pi} \sin(n\phio{}) e^{ix\cos\phio{}} \, d\phio{}$, which
can be evaluated analytically as Bessel functions. Doing so yields the
image of the scattered field at the detector plane as a function of the
sphere's position $\vec{x}_p$ in cylindrical coordinates:
\begin{equation}
\begin{aligned}
\Edet{} (\rhop{}, \phip{}, \zp{}) =
    & \frac 1 2 \frac {E_0} M e^{-ik\radi{}} e^{i\wavefront{}} e^{-ik\rado{}} \, \times \\
    & \left\{ \left[ \Izero{}(k\rhop{}, k\zp{}) + \Itwo{}(k\rhop{}, k\zp{}) \cos(2\phip{}) \right] \unit{x} + \Itwo{}(k\rhop{}, k\zp{}) \sin(2\phip{}) \unit{y} \right\},
\end{aligned}
\label{eqn:spherical_scatterer_lens}
\end{equation}
where we define the integrals $\Izero{}$ and $\Itwo{}$ as
\begin{align}
\Izero{} (u, v) = & \int_0^\lensangle
    \left[ S_\perp(\thetao{}) + S_\parallel(\thetao{})\right] J_0(u \sin \thetao{})
    e^{iv(1- \cos \thetao{}) }
    \sqrt{\cos \thetao{}} \sin \thetao{} \, d\thetao{}
\label{eqn:integral_I0_def}
\\
\Itwo{} (u, v) = & \int_0^\lensangle
    \left[ S_\perp(\thetao{}) - S_\parallel(\thetao{})\right] J_2(u \sin \thetao{})
    e^{iv(1- \cos \thetao{}) }
    \sqrt{\cos \thetao{}} \sin \thetao{} \, d\thetao{},
\label{eqn:integral_I2_def}
\end{align}
where $J_n$ is the Bessel function of the first kind of order $n$.

\subsection{Incident beam and hologram}
\label{sec:hologram}

The change to the incident beam as it passes through the lens is much
simpler to evaluate. As the initially collimated incident beam
propagates to the surface $\So$, it accumulates a phase
$\exp(-ik\rado{})$. Since the incident beam strikes the lens axially,
its polarization vector and amplitude are unchanged, and the
lens only imparts a phase factor $\exp(i\wavefront{})$. On exiting, the
incident beam passes through a focus at the back focal point of the
lens, eventually accumulating a Gouy phase shift of
$-1$~\cite{Mertz2010introduction}. The beam continues to propagate to
the detector plane, accumulating an additional phase shift of
$\exp(-ik\radi{})$. On striking the detector, the incident beam has a
magnification of $M$, which decreases the magnitude of the field by
$1/M$. Combining these effects, we find that the incident field at the
detector plane is
\begin{equation}
\Eindet{} = - \frac {E_0} M e^{-ik\radi{}} e^{i\wavefront{}} e^{-ik\rado{}} \unit{x}.
\label{eqn:incident_field}
\end{equation}
With the addition of the field-rescaling parameter $\alpha$ used to
capture imperfections in the optical setup, the intensity recorded at
the detector is proportional to $|\Eindet{} + \alpha \Edet{}|^2$.
Absorbing the shared phase factors and constants into an overall scaling
factor $\hologram{}_0$ yields an expression for the recorded hologram:
\begin{equation}
\begin{aligned}
\frac {\hologram{}(\rhop{}, \phip{}; \zp{})} {\hologram{}_0} = \,
    & 1 - \alpha \Re\left( \Izero{} (k\rhop{}, k\zp{}) + \Itwo{}(k\rhop{}, k\zp{}) \cos 2\phip{} \right) + \\
    & \frac {\alpha^2} {4} \left| \Izero{}(k\rhop{}, k\zp{}) + \Itwo{}(k\rhop{}, k\zp{}) \cos 2 \phip{} \right|^2 + \frac {\alpha^2} {4} \left| \Itwo{}(k\rhop{}, k\zp{}) \right|^2 \sin^2 2 \phip{}.
\end{aligned}
\label{eqn:mielens_intensity}
\end{equation}

Equation~\eqref{eqn:mielens_intensity} is the basis of what we call the
\emph{lens model}, which predicts the hologram at a single point on the
detector plane for a particle at an arbitrary position $\xp$ from the
conjugate point in the object plane. Calculating a hologram with the
lens model is more straightforward than with the model of
Ref.~\citenum{ovryn2000imaging}, in that it requires the evaluation of
only two integrals $\Izero{}$ and $\Itwo{}$ in
Eqs.~\eqref{eqn:integral_I0_def}-\eqref{eqn:integral_I2_def} and their
combination in Eq.~\eqref{eqn:mielens_intensity}. Although derived for a
point at the center of the detector, Eq.~13 is valid for any point on
the detector plane, owing to the translational invariance of the imaging
system. To calculate the hologram on the entire detector plane, we
repeat this calculation, measuring the particle position $\xp{}$ from
the object point conjugate to each point on the detector plane.

\section{Experimental Results and Discussion}

\begin{figure}
\centering
\includegraphics{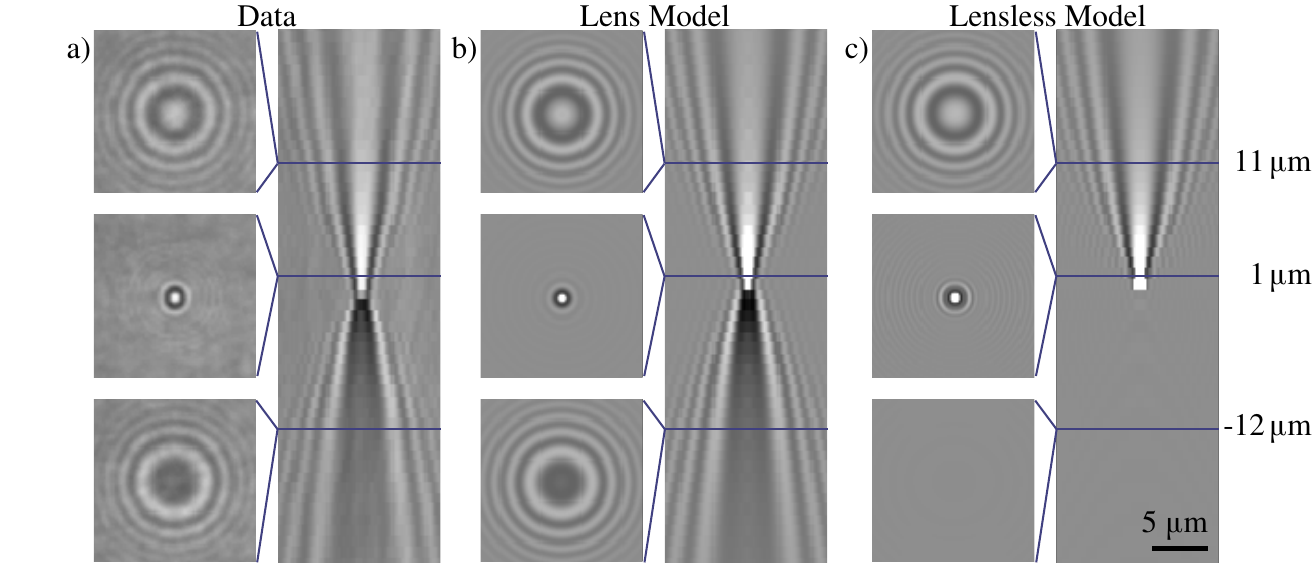}
\caption{ Comparison of (a) experimental holograms of a
  1-\micron{}-diameter polystyrene particle in an acrylamide gel to
  predictions from (b) the lens model and (c) lensless model, over a
  range of particle $\zp{}$ positions. The left half of each panel shows
  holograms for a particle 11~\micron{} above, 1~\micron{} above, and
  12~\micron{} below the focal plane of the objective, while the right
  half of each panel shows $xz$-cross-sections as a function of the
  focal position, where each cross-section is the intensity of the
  hologram across a line running through the center of the central lobe.
  All images are shown at the same scale. The lensless model cannot
  predict the recorded holograms when the particle is in or below the
  focus.}
\label{fig:model_comparison}
\end{figure}

To validate our model, we do an experiment in which we immobilize
1-\micron{}-diameter polystyrene sphere in a gel and obtain holograms of
an isolated particle as we sweep the microscope focus through it (see
Appendix~\ref{appendix:expmethods}). When the particle is far above the focal
plane, the recorded hologram consists of a bright central lobe
surrounded by rings (Fig.~\ref{fig:model_comparison}a). The ring spacing
is primarily set by $\zp{}$ and is visible as the cone-like structures
in the $xz$-cross-section. As the particle approaches the focal plane,
these rings come closer together, until the hologram becomes an image of
the particle in focus. As the particle passes below the focus, the
recorded hologram once again consists of rings centered on a central
lobe with a spacing set primarily by the particle $\zp{}$, but with a
dark rather than bright central lobe.

Both models predict holograms that agree with the experimental data when
the particle is well above the focal plane
(Fig.~\ref{fig:model_comparison}b,c). However, as the particle nears the
focus, the predictions of the lensless model start to deviate from the
measurements. The lensless model is not designed to predict the hologram
when the particle is at or below the focal plane, and in these regions,
the lensless model produces unphysical predictions: it predicts that the
hologram of a particle straddling the focus depends on the particle's
internal field, and that of a particle below the focus depends on the
back-scattered field. In contrast, the lens model uses
only the forward-scattered light and generates holograms that agree with
the experimental data throughout the depth of field.

\begin{figure}
  \centering
  \includegraphics[width=\textwidth]{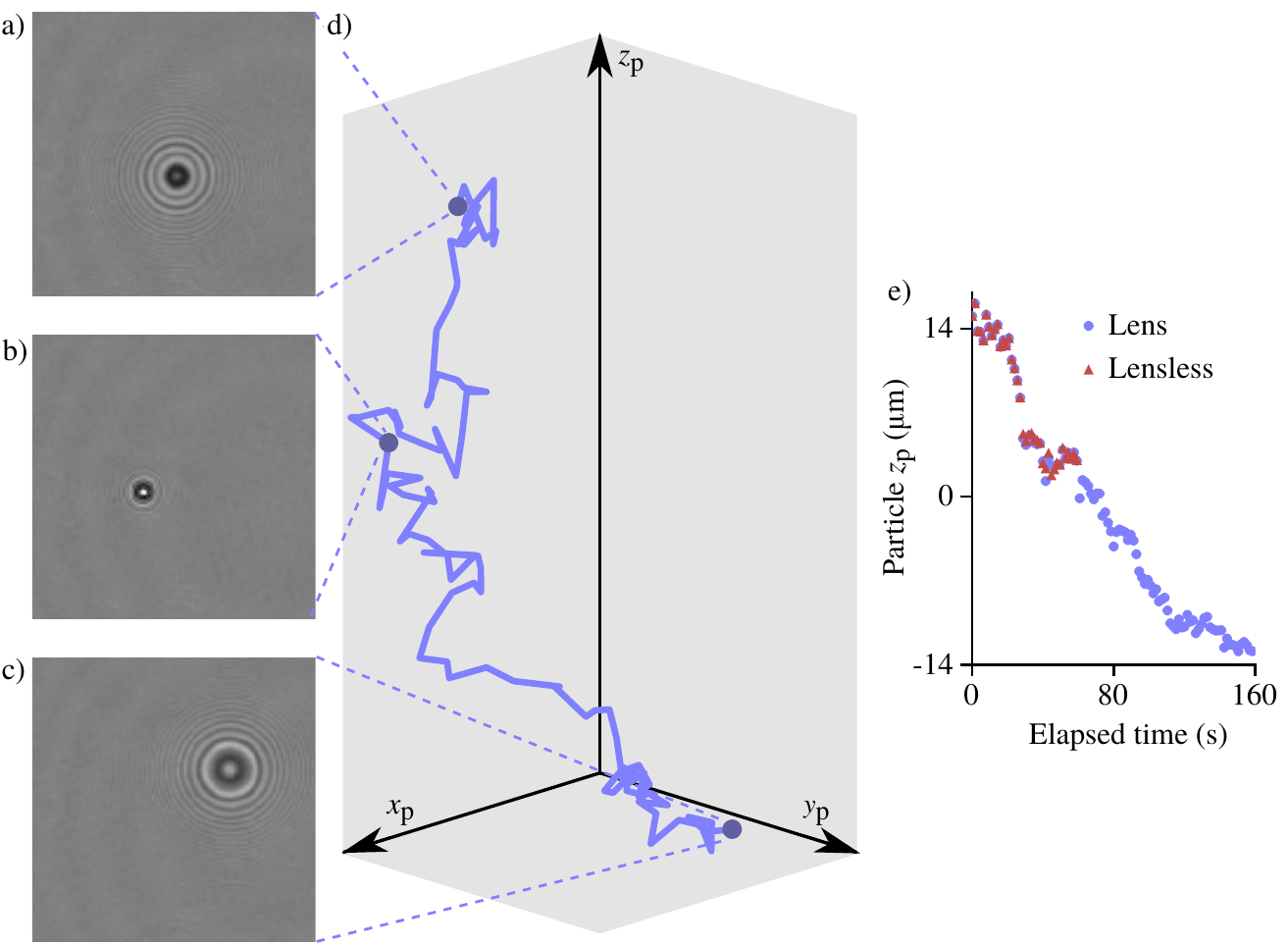}

  \caption{3D tracking of a 2.4-\micron{}-diameter polystyrene sphere as
    it diffuses and sediments through the focus of the objective to the
    coverslip located approximately 15~\micron{} below the focal plane.
    Images show recorded holograms (a) well above the focus, (b) near
    the focus, and (c) well below the focus. From the recorded
    holograms, we reconstruct (d) the particle's 3D trajectory,
    including (e) its height above the focus $\zp{}$ as a function of
    time. Using the lensless model, we track the particle only above the
    focus (red triangles in (e)). By modeling the lens, we extend the
    range of particle tracking to within and below the focus (trajectory
    in (d) and blue circles in (e).
    }

\label{fig:polystyrene-sedimentation}
\end{figure}

To test how the lens model performs in a 3D particle tracking
experiment, we record a movie of a 2.4-\micron{}-diameter polystyrene
particle sedimenting in water. The particle starts tens of micrometers
above the focus and sediments through the focus as it diffuses. We do
not change the focus of the objective during the experiment. We measure
the particle's trajectory by fitting the lens model to each frame of the
video, fitting for the particle's 3D position, radius, and refractive
index, as well as the objective's acceptance angle $\lensangle{}$ and
the field rescaling parameter $\alpha$. For comparison, we also fit the
lensless model to the data when the particle is above the focal plane,
fitting for the particle's 3D position, radius, and refractive index, as
well as the field rescaling parameter $\alpha$. We then compare the
particle trajectories inferred from the two models to each other and to
the sedimentation velocity predicted from Stokes's law
(Fig.~\ref{fig:polystyrene-sedimentation}).

Both above and below the focus, the trajectory inferred using the lens
model is consistent with that expected from Stokes's law for a
diffusing, sedimenting particle. We measure a sedimentation velocity of
$0.18 \pm 0.06$ \micron{}$/$s, where the uncertainty is primarily from
the particle's diffusion. We predict a velocity of $0.18 \pm 0.04$
\micron{}$/$s, where the uncertainty is primarily from limited knowledge
of the particle radius.

When the particle is well above the focus, the positions that we infer
using the lens model agree quantitatively with those we infer using the
lensless model. However, when the particle is within 3~\micron{} of the
focus, the best-fit holograms from the lensless model fail to reproduce
the recorded images. Instead, the best fits from the lensless model
converge to $\alpha = 0$, corresponding to a structureless
hologram.

\begin{figure}
  \centering
  \includegraphics{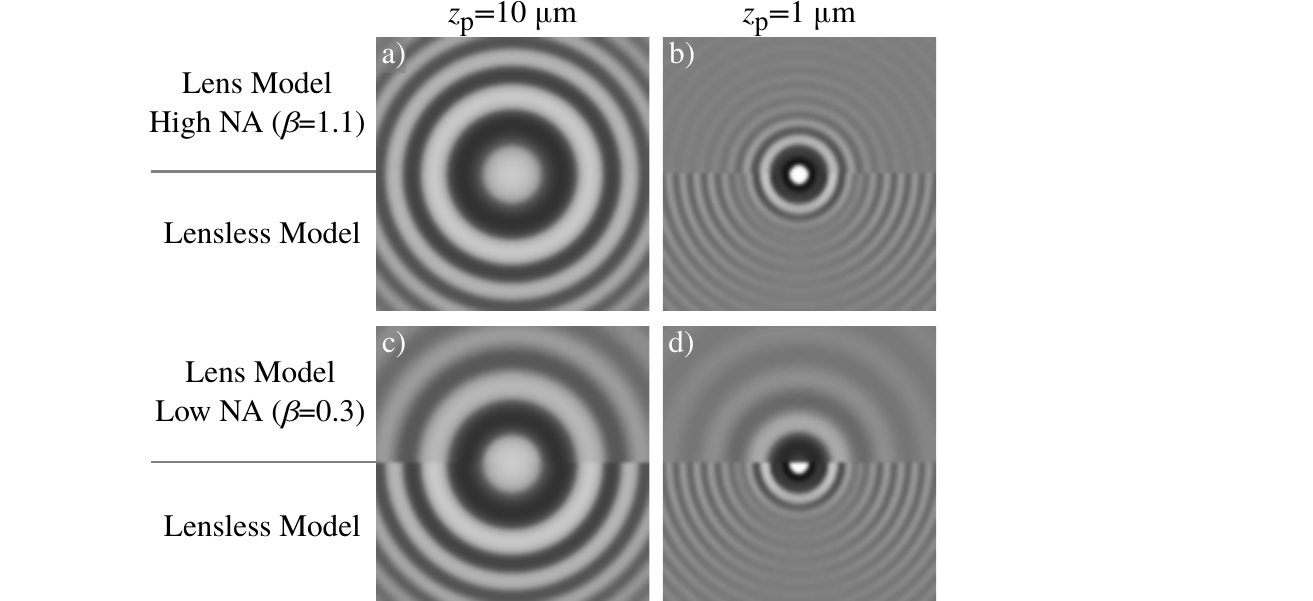}
  \caption{Even above the focus, predictions for the hologram differ
    between the lens and lensless models. We show simulated holograms of
    a 1-\micron{}-diameter polystyrene sphere in water for (a) particle
    height $\zp{}=10$~\micron{} above the focus and lens acceptance
    angle $\lensangle=1.1$; (b) $\zp{}=1$~\micron{} and
    $\lensangle=1.1$; (c) $\zp{}=10$~\micron{} and $\lensangle=0.3$; and
    (d) $\zp{}=1$~\micron{} and $\lensangle=0.3$. The upper half of each
    image shows the hologram simulated with the lens model and the lower
    half shows the simulation from the lensless model. The contrast has
    been nonlinearly adjusted to highlight the fringes, using a gamma of
    $0.5$ for intensities above and below the hologram background
    intensity.
      }
\label{fig:comparemieonlymielens}
\end{figure}

There are differences between holograms predicted by the lensless model
and those predicted by the lens model even when the particle is above
the focal plane. These differences are shown in
Fig.~\ref{fig:comparemieonlymielens}. When the particle is 10~\micron{}
above the focus, a hologram simulated with the lensless model is
indistinguishable from one simulated with the lens model for a
high-$\mathrm{NA}$ lens (Fig.~\ref{fig:comparemieonlymielens}a).
However, as either the $\mathrm{NA}$ decreases
(Fig.~\ref{fig:comparemieonlymielens}b) or the particle approaches the
focus (Fig.~\ref{fig:comparemieonlymielens}c), the two models differ in
their predictions, particularly in the fringe spacing and contrast. When
the particle is close to the focal plane and imaged with a low-aperture
lens, the difference in fringe spacing predicted by the models becomes
dramatic (Fig.~\ref{fig:comparemieonlymielens}d). Furthermore, in this
case the lens model predicts a dark central lobe, while the lensless
model predicts a bright one (Fig.~\ref{fig:comparemieonlymielens}d).

A dark central lobe near the focus is more physically realistic than a
bright one: as the particle passes through the focus, the Gouy phase
shift must lead to an inversion of the contrast of the hologram's
central lobe, as is well appreciated in both brightfield and holographic
microscopy~\cite{Wilson20123d, Mertz2010introduction}. When the particle
is far above the focal plane, the scattered beam comes to a focus behind
the detector and does not have a Gouy phase shift at the detector plane.
But as the particle passes below the focal plane, the scattered beam
comes to a focus in front of the detector, smoothly accumulating a Gouy
phase shift of $\pi$. Thus, the central fringe should become dark near
and below the focus, as predicted by the lens model and as shown in our
measurements (Fig.~\ref{fig:model_comparison}a). The lensless model
predicts no change in contrast as the particle nears the focus, because
it predicts no change in phase of the forward-scattered beam relative to
the incident beam as $\zp{}$ changes.

\begin{figure}
  \centering
  \includegraphics[width=\textwidth]{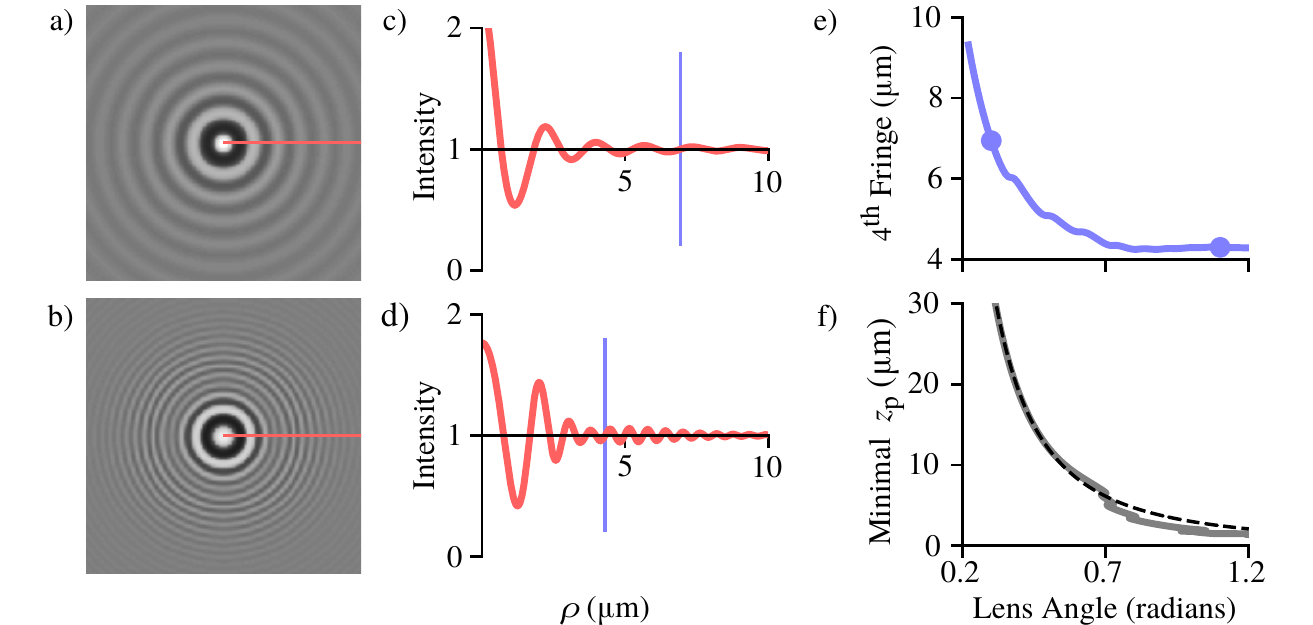}
  \caption{Simulations from the lens model show the effect of lens angle
    on the fringe spacing in a recorded hologram. (a, b) Holograms
    for a 1.0-\micron{}-diameter polystyrene particle
    $\zp{}=5$~\micron{} above the focal plane, simulated from the lens
    model with an acceptance angle of $\lensangle=0.3$ (a) and
    $\lensangle=1.1$ (b). The contrast is nonlinearly adjusted to highlight
    the fringes, as described in Fig.~\ref{fig:comparemieonlymielens}.
    (c, d) The normalized intensity of the holograms in (a) and (b),
    respectively, plotted against distance from the particle center. We
    determine the location of the fourth fringe as the point where the
    intensity crosses the $\rho$-axis immediately before the fourth
    maximum (blue lines). (e) Calculated location of the fourth fringe
    as a function of lens acceptance angle for a 1.0-\micron{}-diameter
    polystyrene sphere located 5~\micron{} above the focal plane. The
    two circles highlight the fringe spacings in panels~c and d. (f)
    Minimum height $\zp{}$ at which the location of the fourth fringe
    predicted by the lensless model is within 2\% of its value in the
    lens model (solid gray line). The dashed black line shows the
    3.0~\micron{} / $\lensangle{}^2$ scaling.
    }
  \label{fig:effectlensangle}
\end{figure}

To understand why the fringe spacing differs between the two models, we
consider the effect of the lens acceptance angle $\lensangle$ on the
fringe spacing. The lens model predicts that near the focus the fringe
spacing should increase significantly with decreasing $\lensangle$, as
shown in Fig.~\ref{fig:effectlensangle}. In the lens model, this
increase in fringe spacing is due to the point-spread function. A lens
images slowly-varying electric fields but blurs out any features with
periodicity less than the objective's resolution limit $\lambda /
\mathrm{NA} = \lambda / n_f \sin \lensangle{}$. As a result, the
smallest possible fringe spacing in a hologram from the lens model is
$\lambda / \mathrm{NA}$, larger than the shortest possible fringe
spacing of $\lambda$ in the lensless model. This difference is
especially noticeable when the fringes are closely spaced, as happens
either far from the hologram's center or when the particle is near the
focus.

The variation of the fringe spacing with lens acceptance angle can lead
to systematic errors when one uses the lensless model to infer the
particle position from a measured hologram. To illustrate this point, we
quantify the change in fringe spacing as a function of the lens
acceptance angle. Specifically, we calculate the distance from the
hologram's central lobe to the fourth fringe
(Fig.~\ref{fig:effectlensangle}a--d) in simulated holograms of a
1.0-\micron{}-diameter polystyrene sphere in water at
$\zp{}=5$~\micron{}. According to the lens model, the distance from the
central lobe to the fourth fringe changes significantly with the lens
acceptance angle $\lensangle{}$, from 6.9~\micron{} when
$\lensangle{}=0.3$ to 4.3~\micron{} when $\lensangle{}=1.1$
(Fig.~\ref{fig:effectlensangle}e). Because information about the
particle height $\zp{}$ is contained primarily in the fringe spacing,
and because the lensless model does not account for the change in fringe
spacing due to the objective's limited resolution, using the lensless
model to infer $\zp{}$ from a measured hologram will lead to a
systematic error.

To determine the imaging conditions under which this systematic error is
significant, we calculate the minimum particle $\zp$ at which a hologram
predicted by the lensless model agrees with that of the lens model
(Fig.~\ref{fig:effectlensangle}f), as quantified by the difference in
the fourth fringe position. As the particle approaches the focus, the
scattered electric field in the focal plane varies on finer and finer
scales. When the particle is closer than a certain $\zp{}$, the field
varies on a finer scale than the objective's resolution limit $\lambda /
\mathrm{NA}$, and the resolution limit rather than $\zp{}$ sets the
fringe spacing in the lens model. At this $\zp{}$, the lens and lensless
models predict different holograms. For a lens with $\lensangle{} = 0.3$
(corresponding to an NA of 0.4 for a water-immersion objective), the
particle $\zp{}$ must be at least 20~\micron{} above the focus for the
two models to agree. As the lens acceptance angle increases to
$\lensangle=0.8$ (NA of 1.0 for water-immersion), the particle can be as
low as 5~\micron{} above the focus before the fringe spacing differs
significantly from that predicted by the lensless model.

The lensless model omits two effects: the Gouy phase shift and the
objective's resolution limit. Both of these effects are important when
the particle is within the objective's depth of field: the Gouy phase
shift because the scattered beam's waist straddles the detector, and the
resolution limit because the hologram's structure varies rapidly. In
fact, we find that the minimal $\zp$ for a hologram to be accurately
described by the lensless model scales as
$3~\micron{}/\lensangle^2$~(Fig.~\ref{fig:effectlensangle}f),
proportional to the theoretical scaling of the objective's depth of
field~\cite{Born1983, Mertz2010introduction}.

Modeling the lens also resolves a puzzle about holographic imaging of
particles near the focus: the lensless model requires the near-field
dependence of the scattered field to predict the hologram
accurately~\cite{Fung2012imaging, Fung2011measuring, Wang2014using},
even though the detector is hundreds of millimeters downstream of the
lens, well in the far field. For a 1~\micron{}-diameter particle
15~\micron{} above the focus, holograms calculated by the lensless model
which use the far-field approximation deviate by 10\% from those which
use the full near-field dependence~\cite{fung2013measuring}. Why does
the near-field form provide a better approximation than the far-field
solution? The reason is that the far-field form of the electric field as
predicted by the lensless model is only an approximate solution to the
wave equation, not an exact solution. In contrast, the near-field form
is an exact solution to the wave equation, albeit for a different
imaging setup than used in holographic microscopy. By constraining the
field to be an exact solution to the wave equation, the near-field
version of the lensless model describes the actual hologram more closely
than the far-field version does. By contrast, the electric field
predicted by the lens model is an exact solution to the wave equation
for the imaging setup.

\section{Conclusion}

We have presented a model that accounts for the effect of an
aberration-free objective lens on in-line holograms of isolated
spherical particles, and we show that the fringe spacing and contrast
depend strongly on the lens acceptance angle. This variation agrees well
with that observed experimentally above, near, and below the focus of
the objective. We have shown also that the lens model can be used to
infer the 3D position of a colloidal sphere near and below the focus of
the objective, effectively doubling the depth of field as compared to
the lensless model. Furthermore, because the model accounts for the
effect of the lens aperture on the fringe spacing, it can be used to
infer the height of the particle from a measured hologram with lower
systematic error, even above the focus.

By capturing physical effects of the lens such as the Gouy phase shift,
our results also provide insight into holograms of more complex
scatterers. The compact representation we derive for homogeneous
spherical scatterers does not apply to more complex scatterers such as
ellipsoids or clusters, but in these cases a lensless model can be used
so long as the scatterer is above the depth of focus of the objective.
The lens model can be used to quantitatively estimate this minimum
height as a function of the numerical aperture of the objective.
Therefore, the model is useful not only for fitting holograms of
spherical scatterers, but also for determining the range of validity of
general lensless models, which might be easier to implement
computationally.

For spherical particles, all the physics described by the lens model is
also described in the earlier work of Ref.~\citenum{ovryn2000imaging}.
However, our simplified model requires the calculation of only the
integrals $\Izero{}$ and $\Itwo{}$ in Eqs.~\eqref{eqn:integral_I0_def}
and \eqref{eqn:integral_I2_def}, allowing for a straightforward and
efficient computational implementation. We find that calculating a
hologram with the lens model takes roughly the same time as with the
lensless model, because much of the time is spent computing the Mie
scattering matrices. With the numerical optimizations detailed in
Appendix~\ref{appendix:nummethods}, the implementation of the lens model
becomes three times as fast as the lensless model. This fast
implementation is available in the open-source package
\texttt{holopy}~\cite{barkley2019holographic}.

\appendix
\section{Numerical Methods}
\label{appendix:nummethods}

Calculating a hologram of shape $N \times N$ with the lens model for a
single particle involves numerically integrating
Eqs.~\eqref{eqn:integral_I0_def}--\eqref{eqn:integral_I2_def} over each
of the $N^2$ values of $\rho_p$ in the image. We perform this
integration through Gauss-Legendre quadrature over $\cos
\theta_\mathrm{o}$. For large $\rho_p$, the $J_n(u
\sin(\theta_\mathrm{o}))$ terms oscillate rapidly and require many
quadrature points for accurate integration; empirically, we find 100
quadrature points provides good accuracy for $k\rho_p < 400$. Because
the Mie scattering coefficients $S_\parallel(\theta)$ and
$S_\perp(\theta)$ do not depend on $\rhop$ or $\zp$, we pre-compute the
scattering coefficients once at the quadrature nodes and use these
values for each of the separate integrations over $\rho_p$. With the
pre-computed scattering coefficients, the slowest part of the
integration is the computation of the Bessel functions at the $N^2$
values of $\rhop$. Since the maximum $\rhop$ is only of $O(N)$, we
reduce this near-redundant computation by creating an interpolator for
$\mathcal{I}_n$. Interpolating $\mathcal{I}_n$ with piecewise Chebyshev
interpolators of degree 32 over windows of size $\Delta u = 39$ provides
a fifteen-fold speed increase, with a relative accuracy on the order of
$10^{-12}$. We use \texttt{holopy} to calculate the lensless model,
using methods described in~\cite{barkley2019holographic}.

We validate this numerical implementation of
Eqs.~\eqref{eqn:spherical_scatterer_lens}--\eqref{eqn:mielens_intensity}
with the following test derived from using conservation of energy of the
scattered beam. The power scattered from the particle crossing the
surface $\So{}$ must be equal to that striking the detector plane. By
construction, the Poynting vector is perpendicular to $\So$, so the flux
on $\So$ is $|\vec{E}_{\So}|^2/c$. Substituting the field from a Mie
scatterer (Eqs.~\eqref{eqn:farfield_so} and~\eqref{eqn:miescattering})
and integrating over $\phio$ gives the power on $\So{}$:
\begin{equation}
P_{\So} = \frac \pi c \frac {|E_0|^2} {k^2} \int_0^\lensangle
    \left[ |S_\parallel(\theta)|^2 + |S_\perp(\theta)|^2 \right]
    \sin \thetao d\thetao.
\label{eqn:power_scattered_entrance}
\end{equation}
To the accuracy of the approximations, the Poynting vector is
perpendicular to the detector plane, and the flux on the detector is
$|\Edet{}|^2 / c$. Substituting the scattered field from
Eq.~\eqref{eqn:spherical_scatterer_lens} and integrating over $\phid$
gives
\begin{equation}
P_\mathrm{det} = \frac \pi {2c} \frac {|E_0|^2} {M^2} \int_0^\infty
    \left[ |\Izero{}(k\rhod{} / M; k\zp)|^2 + |\Itwo{}(k\rhod{} / M; k\zp)|^2 \right] \rhod{} \, d\rhod{}
\label{eqn:power_scattered_detector}
\end{equation}
Equating the powers in Eqs.~\eqref{eqn:power_scattered_entrance}
and~\eqref{eqn:power_scattered_detector} and substituting $u \equiv
k\rhod{} / M$ in Eq.~\eqref{eqn:power_scattered_detector} relates the
scattered field on the detector plane to the scattering matrix as
\begin{equation}
\int_0^\lensangle
    \left[ |S_\parallel(\theta)|^2 + |S_\perp(\theta)^2| \right]
    \sin \theta\, d\theta
    =
\frac 1 2 \int_0^\infty
    \left[ |\Izero{}(u, k\zp)|^2 + |\Itwo{}(u, k\zp)|^2 \right] u\, du.
\label{eqn:homodyne_power_check}
\end{equation}
We use this integral relationship as a unit test of our numerical
implementation of the scattered field, testing the implementation over
a range of lens acceptance angles, particle positions, particle
refractive indices, and particle radii. The analogous test for
conservation of total energy in the scattered and incident beams is not
numerically practical; however, one can show that conservation of energy
combined with Eqs.~\eqref{eqn:mielens_intensity} yields the optical theorem.

\section{Experimental Methods}
\label{appendix:expmethods}

The data in Fig.~\ref{fig:model_comparison} were obtained on a sample of
colloidal spheres in a polyacrylamide hydrogel prepared by free radical
polymerization. To make the hydrogel, we first combine 2 parts water and
1 part Protogel (30\% w/v acrylamide, 0.8\% w/v bis-acryl-amide,
National Diagnostics). We then mix 27~\microliter{} of the Protogel
mixture with 1.5~\microliter{} of a $10^{-4}$\% w/v suspension of
1.0-\micron{}-diameter polystyrene spheres (Polysciences 19404) in water
and with 0.2~\microliter{} of the UV polymerization initiator Darocur
1173 (2-hydroxy-2-methylpropriophenone, Aldrich). We place the mixture
in a sample cell composed of glass slides separated by a 50~\micron{}
thick spacer, seal the edges of the cell with vacuum grease, and
irradiate the cell with an ultraviolet lamp to cure the hydrogel. This
procedure produces a transparent polyacrylamide hydrogel with refractive
index of 1.348, as measured by an Abbe refractometer. To prevent the
hydrogel from drying during the experiment, we fill the chamber
surrounding the gel with water using a syringe inserted through the
vacuum grease. We then image the particle at 51 defocus positions
spanning 50~\micron{}, centered on the particle. Details of the imaging
setup are given below.

For the sedimentation experiment in
Fig.~\ref{fig:polystyrene-sedimentation}, we image a dilute suspension
of polystyrene microspheres with a reported diameter of 2.4~\micron{}
(density 1.055 g/mL, index 1.591 at 590 nm, Invitrogen S37502). We
prepare a $10^{-4}$ \% w/v colloidal suspension and place it in a sample
chamber made of a glass coverslips separated by 50-\micron{}-thick
plastic spacers. We invert the sample chamber for several minutes to
give the particles time to sediment to the top coverslip, then place the
sample chamber right-side up onto the microscope. We adjust the focal
position of the objective to be within the sample chamber and image the
particles as they sediment through the focus.

In both experiments, the imaging setup consists of a Nikon Eclipse Ti
TE2000 microscope with a water-immersion objective (Plan Apo VC
60$\times$/1.20 WI, Nikon), with the correction collar set to the
thickness of the coverslip to minimize aberrations. The images in
Fig.~\ref{fig:model_comparison} are obtained from a
1280$\times$1024-pixel CMOS color sensor array (Edmund Optics 1312C) and
in Fig.~\ref{fig:polystyrene-sedimentation} from a
1024$\times$1024-pixel CMOS monochrome sensor array (PhotonFocus A1024).
After obtaining images of each particle, we record a set of background
images by moving the particle out of the field of view in several
directions. The raw holograms are processed prior to fitting by dividing
by the average background intensity. In addition, we take dark-count
data by acquiring images with the primary light source switched off. We
average these dark-count images and subtract them from the
background-corrected data before analysis. For the sedimentation data,
we fit the processed holograms using a
parallel-tempered~\cite{earl2005parallel},
affine-invariant~\cite{goodman2010ensemble}, Markov-chain Monte Carlo
ensemble sampler, as implemented in the Python package
\texttt{emcee}~\cite{foreman2013emcee}. We find that parallel tempering
improves tracking by avoiding local minima in the posterior landscape.

\section*{Funding}

We acknowledge support from the National Science Foundation
through grant number DMR-1420570.

\section*{Acknowledgments}
We thank G. Keshavarzi, A. Goldfain, R. Perry, T. Dimiduk, J.
Fung, and A. Small for useful discussions.

\section*{Conflict of interest statement}

The authors declare no conflicts of interest.

\end{document}